\documentclass[lettersize,journal]{IEEEtran}
\usepackage{amsmath,amsfonts}
\usepackage{algorithmic}
\usepackage{algorithm}
\usepackage{array}
\usepackage[caption=false,font=normalsize,labelfont=sf,textfont=sf]{subfig}
\usepackage{textcomp}
\usepackage{stfloats}
\usepackage{url}
\usepackage{verbatim}
\usepackage{graphicx}
\usepackage{cite}
\usepackage{bm}
\begin{document}

\title{Generative Joint Source-Channel Coding for Semantic Image Transmission}

\author{Ecenaz Erdemir, ~\IEEEmembership{Member,~IEEE}, Tze-Yang Tung, ~\IEEEmembership{Graduate Student Member,~IEEE}, \\ Pier Luigi Dragotti, ~\IEEEmembership{Fellow,~IEEE}, and Deniz G\"{u}nd\"{u}z, ~\IEEEmembership{Fellow,~IEEE}
        % <-this % stops a space
\thanks{E. Erdemir, T.-Y. Tung, P. L. Dragotti, and D. G\"{u}nd\"{u}z are with the Department of Electrical and Electronic Engineering, Imperial College London, London SW7 2AZ, U.K. D. G\"{u}nd\"{u}z is also with the Theory Lab, Central Research Institute, 2012 Labs, Huawei Technologies Co., Ltd., Hong Kong SAR, China (e-mails: (e.erdemir17, tze-yang.tung14, d.gunduz)@imperial.ac.uk) }% <-this % stops a space
}



\maketitle

\begin{abstract}
Recent works have shown that joint source-channel coding (JSCC) schemes using deep neural networks (DNNs), called DeepJSCC, provide promising results in wireless image transmission. However, these methods mostly focus on the distortion of the reconstructed signals with respect to the input image, rather than their perception by humans. However, focusing on traditional distortion metrics alone does not necessarily result in high perceptual quality, especially in extreme physical conditions, such as very low bandwidth compression ratio (BCR) and low signal-to-noise ratio (SNR) regimes. In this work, we propose two novel JSCC schemes that leverage the perceptual quality of deep generative models (DGMs) for wireless image transmission, namely InverseJSCC and GenerativeJSCC. While the former is an inverse problem approach to DeepJSCC, the latter is an end-to-end optimized JSCC scheme. In both, we optimize a weighted sum of mean squared error (MSE) and learned perceptual image patch similarity (LPIPS) losses, which capture more semantic similarities than other distortion metrics. 
InverseJSCC performs denoising on the distorted reconstructions of a DeepJSCC model by solving an inverse optimization problem using style-based generative adversarial network (StyleGAN). Our simulation results show that InverseJSCC significantly improves the state-of-the-art (SotA) DeepJSCC in terms of perceptual quality in edge cases. In GenerativeJSCC, we carry out end-to-end training of an encoder and a StyleGAN-based decoder, and show that GenerativeJSCC significantly outperforms DeepJSCC both in terms of distortion and perceptual quality.
\end{abstract}


\section{Introduction}
\IEEEPARstart{C}{ommunication} systems are designed and optimized to reliably transmit information from one point to another over noisy communication channels. They consist of three blocks: an encoder at the transmitter, a noisy channel, and a decoder at the receiver. From the first to the fifth generation of communication systems, the encoding process has followed the conventional two-step approach of Shannon's separation theorem \cite{shannon1948mathematical}, which decomposes the transmitter into a source encoder and a channel encoder. The former removes the redundant information from the source to allow reconstruction at the desired quality, while the latter protects the compressed information against errors introduced over the wireless channel. Similarly, the receiver consists of source and channel decoders. Shannon proved that separate source and channel coding provides theoretical optimality in the case of infinitely long code blocks. On the other hand, in practical systems that require source transmission under extreme latency, bandwidth, or energy constraints, it is known that the separation approach can be highly suboptimal. Such constraints apply in most emerging applications, such as Internet-of-everything (IoE), vehicle-to-everything (V2X) communications, eHealth services, and augmented/virtual reality (AR/VR).

Although joint source-channel coding (JSCC) has been known to achieve better performance than separate source compression followed by channel coding, practical design of such joint coding schemes has been a long-standing challenge. Recently, deep learning (DL) based JSCC methods, e.g., DeepJSCC, have shown outstanding results due to their ability to extract complex features from the training data while incorporating the channel characteristics into their encoding implicitly \cite{bourtsoulatze2019deep,burth2020joint,kurka2020deepjscc,tung2022deepjscc,tung2021deepwive,yang2022ofdm,kurka2021bandwidth,xu2021wireless,choi2019neural}. However, these DL-based JSCC methods do not focus on the semantic similarities between the source signal and its reconstruction at the receiver and have mostly considered the mean squared error (MSE) or structured similarity metrics (SSIM/MS-SSIM) as the end-to-end measure of distortion, and the loss function for training. On the other hand, in semantic communications \cite{gunduz2022beyond}, particularly in the context of image/video transmission, the receiver is not necessarily interested in reconstructing the original source signal with minimal distortion. Instead, the receiver may be interested in some downstream task, such as classification \cite{jankowski2020joint,lee2019deep}, or retrieval \cite{jankowski2020wireless}, in which case it would be sufficient to only convey the relevant features of the source signal for the prescribed task. Alternatively, for extreme image compression \cite{agustsson2019generative}, the receiver may be interested in generating an output with the same distribution as the source signal rather than its accurate representation. In image compression literature, this aspect has recently been acknowledged as the perception loss, resulting in the rate-distortion-perception trade-off \cite{blau2019rethinking}. In general, the perception loss is measured by the divergence between the source and reconstruction distributions and can be minimized in practice using generative adversarial networks (GANs) \cite{tschannen2018deep}.

In this work, we propose two JSCC schemes for wireless image transmission that are designed and optimized to tackle the extreme physical limitations of the environment, while reconstructing an output image that minimizes the desired fidelity measure with respect to the particular input image and looks realistic to human perception. Some examples of the \textit{edge cases} that can negatively affect the communication overhead are very low bandwidth compression ratio (BCR) and very low channel signal-to-noise ratio (SNR). In classical DeepJSCC, the goal is to reconstruct the source signal, e.g., image, at the receiver with minimum distortion by jointly optimizing the source and the channel coding via deep neural networks (DNNs). However, this often results in a significant loss of perceptual quality for the edge cases. We overcome this by following a context-aware communication approach that utilizes style-based generative adversarial network (StyleGAN) generators as well as distortion metrics that align well with the human perception in our optimizations. 

Past works on DeepJSCC train the encoder/decoder networks on patches of images from a large dataset, which are then used and evaluated for the transmission of high-resolution images from the Kodak dataset \cite{bourtsoulatze2019deep,burth2020joint,kurka2020deepjscc,tung2022deepjscc,tung2021deepwive,yang2022ofdm,kurka2021bandwidth,xu2021wireless}. It has been observed that the larger and richer the training dataset the better the performance. On the other hand, the performance improves further if the statistics of the training dataset match that of the test images. This is shown in \cite{burth2020joint} by training and testing DeepJSCC on satellite images that exhibit a particular statistic, where the gain from DeepJSCC with respect to conventional separation-based approaches becomes even more significant. In our first scenario in this work, we consider a DeepJSCC encoder/decoder pair trained on a generic training dataset, e.g., ImageNet. Later, we are given images from another dataset to be transmitted wirelessly. We assume that the encoder network remains the same, but the decoder can deduce the knowledge of the dataset to improve the quality of the reconstructed images. We propose a novel inverse problem approach for this scenario, called \textit{InverseJSCC}. InverseJSCC performs denoising on the distorted reconstructions of DeepJSCC by solving an unsupervised inverse problem and recovers a high-quality source image with better perceptual quality. Exploiting the learned distribution for a particular dataset, e.g., face images, we use a DeepJSCC model and a pre-trained StyleGAN-2 \cite{Karras2019stylegan2} generator, whose architecture is the SotA for high-quality image generation, to generate new face images that perceptually match the source image. While InverseJSCC significantly improves classical DeepJSCC perceptually, it also allows using any differentiable encoder/decoder pair to reconstruct the noisy images. We also show that, unlike DeepJSCC, InverseJSCC does not lose significant performance when the statistics of the training dataset do not match that of the test images. 

Next, we consider end-to-end training of the encoder and decoder using the learned distribution. The proposed solution is called \textit{GenerativeJSCC}, which is composed of an encoder, a non-trainable channel, and a decoder that incorporates residual DNNs and the StyleGAN-2 generator structure. We perform end-to-end training of the encoder/decoder pair of GenerativeJSCC by optimizing a distortion measure that also takes perceptual quality into account.
Our results show that GenerativeJSCC outperforms classical DeepJSCC in terms of both traditional distortion metrics (e.g., mean-squared error) and perceptual similarity metrics that align with human perception well.

\subsection{Our Contributions}
We can summarize the main contributions of this work as
follows:
\begin{itemize}
    \item We study the wireless image transmission problem by considering both the end-to-end distortion and the perceptual quality of the reconstructed image. To achieve realistic high-quality image reconstruction, we employ a pre-trained generator at the receiver and treat the problem as an unsupervised image reconstruction problem.
    \item We first propose InverseJSCC, which uses a pre-trained GAN generator to invert the wireless communication problem in an unsupervised manner and improves the perceptual quality of the state-of-the-art (SotA) DeepJSCC by exploiting the learned distribution of the transmitted images at the receiver. To the best of our knowledge, this is the first unsupervised inverse problem approach to the wireless image communication problem.
    \item We then propose GenerativeJSCC, which is a supervised end-to-end solution to the wireless image communication problem using a pre-trained GAN. 
    \item We carry out extensive simulations using ImageNet and CelebA-HQ datasets and show that InverseJSCC can help improve the perceptual quality of the reconstructed images even though the encoder remains the same. On the other hand, GenerativeJSCC significantly outperforms the SotA DeepJSCC model in edge cases in terms of both distortion and perception quality.
\end{itemize}

\section{Related Work}
DL techniques for JSCC and its various wireless communication applications have attracted a lot of attention in the past decade \cite{bourtsoulatze2019deep,burth2020joint,kurka2020deepjscc,tung2022deepjscc,tung2021deepwive,yang2022ofdm,kurka2021bandwidth,xu2021wireless}.
The first work employing DNNs in wireless image transmission, called \textit{DeepJSCC}, was proposed in \cite{bourtsoulatze2019deep}. The authors used an autoencoder architecture to represent a wireless communication system that is composed of an encoder, a non-trainable channel and a decoder. The encoder/decoder pair is jointly trained such that the encoder learns a function that maps the source image directly to continuous channel inputs and the decoder learns to reconstruct the image from its noisy observations. They demonstrated in \cite{bourtsoulatze2019deep} that DeepJSCC outperforms classical separation schemes with compression \cite{christopoulos2000JPEG} and channel coding \cite{gallager1963LDPC}, and it shows success in adapting to different source or channel types and avoiding the \textit{cliff effect}. DeepJSCC has later been extended to different channel models \cite{kurka2020deepjscc,yang2022ofdm} as well as to video transmission \cite{tung2021deepwive}. However, previous JSCC approaches mostly focus on the rate-distortion trade-off in terms of classical pixel-wise distortion metrics, and often disregard human perception.
In \cite{yang2022ofdm}, the authors study JSCC combined with orthogonal frequency division multiplexing (OFDM) to cope with multi-path fading. The authors adopt a GAN formulation to combine the MSE loss with the adversarial loss of a discriminator network, which distinguishes whether the image is original or generated by the decoder. The generator network of \cite{yang2022ofdm} is jointly trained with the encoder by following the traditional adversarial training approach. However, GAN-generated image quality relies heavily on the range of the generator. The generator network of \cite{yang2022ofdm} cannot achieve the perceptual image quality of the SotA StyleGAN generator, which follows the methodology of progressive growing GANs \cite{karras2017progressive}, and hence produces remarkably high-quality images. Moreover, it has been shown that the traditional distortion metrics, e.g., MSE, PSNR, and SSIM, also used in \cite{yang2022ofdm}, are not as successful in representing human judgments compared to the learned perceptual image patch similarity (LPIPS) metric \cite{zhang2018unreasonable}.

Deep generative models (DGMs) have recently shown immense perceptual quality, which refers to being perceived by humans as a valid (natural) sample \cite{blau2019rethinking}. Therefore, another line of DNN-based JSCC work has utilized well-known DGMs, such as Variational Autoencoders (VAEs), GANs, and etc., in wireless image transmission \cite{choi2019neural,saidutta2020vae,marchioro2020adversarial,erdemir2022privacy}. The idea behind these JSCC approaches comes from the similarity between the wireless communication systems and inverse problems, in which an unknown signal, image or a multi-dimensional volume is reconstructed from its observations \cite{ongie2020deep}. In this analogy, the observations are obtained from a forward process, e.g., an encoder and a noisy channel, which are then inverted by a decoder in a supervised manner to reconstruct the original signal. Hence, a wireless communication system is optimized to learn an identity function such that the reconstructed message matches the source message. This supervised inverse problem approach is similar to denoising autoencoders \cite{mousavi2015denoisingae}, which was incorporated into the wireless communication setting in \cite{choi2019neural}. The authors propose a VAE-based JSCC scheme for a binary symmetric channel and a binary erasure channel system, later extended to wiretap channel scenario by \cite{erdemir2022privacy}. Despite their exceptional generative capability, GANs have only been considered in a JSCC scheme in the context of secure image transmission \cite{marchioro2020adversarial}, where generative adversarial training is used for image reconstruction at the receiver. Despite utilizing DGMs, these approaches still do not incorporate the SotA DGM structures into the wireless communication problem in order to reconstruct high perceptual quality images.

Unsupervised inverse problems, on the other hand, do not rely on a matched dataset of the input images and measurements, since the input images are not available at test time \cite{ongie2020deep}. Instead, they estimate the best potential input image that goes through the known forward process and matches with the observed measurements. However, prior assumptions on the input image distribution, e.g., sparsity assumption in classical compressed sensing \cite{donoho2006compressed} are needed. Pre-trained DGM-based approaches have recently received great attention due to their immense capacity of input data representation \cite{bora2017compressed,menon2020pulse,daras2021intermediate}. The first DGM-based inverse problem approach, called CSGM, was proposed in \cite{bora2017compressed}. The authors solved the classical compressed sensing problem, i.e., $Ax+\text{noise}=y$, by assuming that the input image $x$ is in the range of a pre-trained DGM. Their results show that CSGM outperforms traditional compressed sensing solutions for denoising, inpainting, and compressed sensing, when the input image is in the range of a pre-trained VAE or GAN generator. Later, \cite{menon2020pulse} improved CSGM by including a regularization term that limits the latent input of the generator network to a spherical ball. PULSE \cite{menon2020pulse} also used a pre-trained StyleGAN-2 generator \cite{Karras2019stylegan2}. 
Another work that uses GAN priors for inverse problems is proposed in \cite{daras2021intermediate}, namely intermediate layer optimization (ILO). As its name suggests, ILO solves the classical compressed sensing objective by optimizing the intermediate layers of a pre-trained StyleGAN-2 generator. The algorithm first applies CSGM to find the best initialization for the latent input of the GAN generator, then the objective is optimized by solving projected gradient descent for each intermediate layer recursively. ILO \cite{daras2021intermediate} significantly outperforms the previous approaches, hence it is the SotA inverse problem solution that uses GAN priors.

To the best of our knowledge, unsupervised inverse problems, which show exceptional denoising performance, have not been studied in the context of wireless communication. For both supervised and unsupervised methods, there is a demand in the field to incorporate the generative capabilities of SotA DGMs into wireless image transmission problems.

\section{Problem Statement}
\label{sec:problem_statement}
We consider a communication scenario in which a user wants to reliably transmit a source signal from one point to another over a noisy channel. Our goal is to maximize the semantic similarity between the source and the reconstructed signals at the receiver while minimizing the distortion caused by the channel. We measure the semantic similarity using the perceptual quality metric LPIPS introduced in \cite{zhang2018unreasonable}. In \cite{zhang2018unreasonable}, the authors argue that LPIPS is a distortion measure that better aligns with human perception than the typical pixel-wise metrics. 
We consider a transmitter that maps the source vector $\mathbf{x} \in \mathbb{R}^m$ into a vector of complex-valued channel input symbols $\mathbf{z} \in \mathbb{C}^k$ by an encoding function $f: \mathbb{R}^m \rightarrow \mathbb{C}^k$, where $m$ and $k$ are source and channel bandwidth, respectively. We define the BCR as
\begin{equation}
    \rho=\frac{k}{m},
\end{equation}
which represents the level of compression applied to the source signal. In the case of an image source, the input size would be $m=H\times W \times C$, where $H, W$, and $C$ are the height, width, and color dimensions of the source image.
An average transmit power constraint $\bar{P}$ is imposed at the output of the encoder before the signal is transmitted through the channel, such that
\begin{equation}
\frac{1}{k}\mathbb{E}_{\mathbf{z}}[\|\mathbf{z}\|_2^2]\leq \bar{P},
\end{equation}
where the expectation is over the distribution of the encoded signal. 
The power constraint is implemented at the immediate output of the encoder by normalizing the encoded signal according to: 
\begin{equation}
\mathbf{z}=\sqrt{k\bar{P}}\frac{\mathbf{\tilde{z}}}{\sqrt{\mathbf{\tilde{z}}^H\mathbf{\tilde{z}}}},
    \label{eq:power_normalization}
\end{equation}
where $\mathbf{\tilde{z}}$ is the input to the final normalization at the output of the encoder, and $H$ refers to the Hermitian transpose.
The normalized encoded signal $\mathbf{z}$ is transmitted over the noisy channel that applies a random corruption function $\eta: \mathbb{C}^k \rightarrow \mathbb{C}^k$ that turns $\mathbf{z}$ into the corrupted signal $\mathbf{\hat{z}}$. We consider the widely used additive white Gaussian noise (AWGN) channel model and assume it is known by both the transmitter and the receiver. The transfer function of the AWGN channel is

\begin{equation}
\eta(\mathbf{z},\sigma^2)=\mathbf{z}+\mathbf{n}_C,
\label{eq:channel}
\end{equation}
where the channel noise vector $\mathbf{n}_C$ is sampled in an independent identically distributed (i.i.d.) manner from a circularly symmetric complex Gaussian distribution, i.e., $\mathbf{n}_C\sim \mathcal{CN}(0,\sigma^2 I_{k \times k})$, and $\sigma^2$ is the channel noise power known by both the transmitter and the receiver. Accordingly, the channel SNR is defined as
\begin{equation}
    \text{SNR} = 10 \log_{10} \frac{\bar{P}}{\sigma^2} \  \text{dB}.
\end{equation}
The noisy channel output is observed by the receiver, which then decodes it to an approximate reconstruction $\mathbf{\hat{x}} \in \mathbb{R}^m$ of the source signal by a decoding function $g: \mathbb{C}^k \rightarrow \mathbb{R}^m$.

\begin{figure}[!t]
\centering
\includegraphics[width=3.4in]{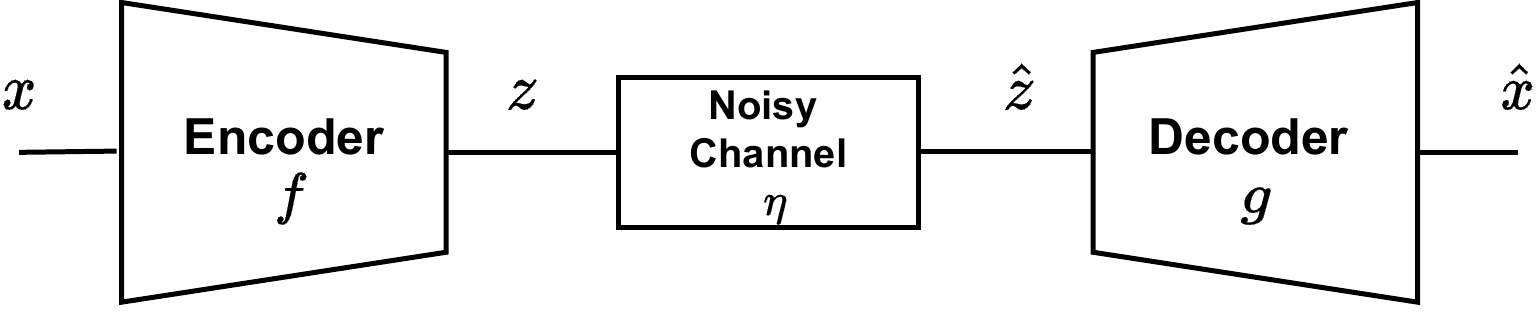}
\caption{Communication system with joint source-channel coding.}
\label{fig:djscc}
\end{figure}

The entire pipeline is optimized by jointly designing the encoder and the decoder functions such that the average distortion between the original source $\mathbf{x}$ and its reconstruction $\mathbf{\hat{x}}$ at the output of the receiver is minimum:  
\begin{equation}
    \min \limits_{f,g} \mathbb{E}_{p(\mathbf{x},\mathbf{\hat{x}})}\big[d(\mathbf{x},\mathbf{\hat{x}})\big],
    \label{eq:rate-distortion}
\end{equation}
where $p(\mathbf{x},\mathbf{\hat{x}})$ is the joint distribution of the source and its reconstruction, and $d(\mathbf{x},\mathbf{\hat{x}})$ can be any measure. In a typical communication scheme, compression schemes usually try to minimize the distortion at any given bit rate, for instance, by minimizing the MSE or maximizing the PSNR, SSIM, etc. However, typical distortion measures do not necessarily imply high perceptual quality; in fact, minimizing the distortion might reduce the perceptual quality \cite{blau2019rethinking}. Therefore, throughout this paper, we focus on a distortion measure that takes both perceptual and pixel-wise similarities into account.
Fig. \ref{fig:djscc} illustrates a simplified diagram of the communication system considered herein. 

In the following section, we propose two JSCC schemes that effectively minimize the distortion between the source signal and its reconstruction, while maximizing perceptual generation quality with the help of DGMs. 
Inspired by the SotA DL-based JSCC approaches \cite{bourtsoulatze2019deep,kurka2020deepjscc,tung2022deepjscc,tung2021deepwive}, we consider DNN-based encoder/decoder pairs that also benefit from the generative capabilities of DGMs in both approaches.
The first scheme can be considered as an inverse problem approach to the optimization (\ref{eq:rate-distortion}), in which the encoder/decoder functions $f$ and $g$ are fixed and a high-quality approximation of $\mathbf{x}$ is recovered from reconstructions $\mathbf{\hat{x}}$ using DGMs. In the second scheme, given $\mathbf{x}$, Eqn. (\ref{eq:rate-distortion}) is minimized by jointly optimizing $f$ and $g$ in a supervised manner.

\section{Proposed Solutions}

Recently proposed CNN-based approaches to JSCC \cite{bourtsoulatze2019deep,burth2020joint,kurka2020deepjscc,tung2022deepjscc,tung2021deepwive} are based on the autoencoder architecture and have focused on the end-to-end distortion between the source and the reconstructed image. The encoder and decoder functions, $f$ and $g$, are modeled as DNNs, which are jointly trained. The goal of the encoder is to transform the source image to a form that can be conveyed through the channel; that is, it satisfies the power and bandwidth constraints. The decoder aims at reversing the impact of the encoder as well as the channel on the input image. However, these works did not consider the perceptual quality of the reconstructed images. 

A GAN-based distortion measure is considered in \cite{yang2022ofdm}, but their decoder acts as the generator directly, which is trained as part of the encoder/decoder pair. In contrast, our work takes advantage of the outstanding generative performance of pre-trained SotA GANs as well as distortion metrics that align with human visual judgment in order to obtain high perceptual quality reconstructions. Firstly, in InverseJSCC, we improve the classical DeepJSCC model by inverting the communication pipeline with the help of a GAN generator by exploiting the learned statistics of the source images at the receiver. This requires solving an inverse problem, in which the forward operator to be inverted is represented by the encoder-channel-decoder architecture from \cite{tung2021deepwive}. In a sense, we optimize the input of the generator model to produce a certain image; that is, we find a latent vector, whose output transmitted by the DeepJSCC encoder through the channel and reconstructed by the DeepJSCC decoder, gives the observed reconstruction. 
InverseJSCC significantly improves the performance of SotA DeepJSCC schemes perceptually in cases where the training dataset statistics of the encoder/decoder pair match with the test dataset, as well as those that do not.
On the other hand, in GenerativeJSCC, we propose an end-to-end training scheme for JSCC which consists of an encoder, a non-trainable channel, and a GAN-based decoder.

\subsection{Inverse Problem Approach to Semantic Communications}
\label{sec:inverse_problems}

In this section, we introduce InverseJSCC as an unsupervised improvement on the DeepJSCC approach by treating the end-to-end reconstruction by DeepJSCC as the FP, and model the reconstruction of the source image with better perceptual quality as an \textit{inverse problem}. More specifically, InverseJSCC treats the reconstructions at the receiver as the \textit{measurements}, obtained by an unknown data sample going through a typically non-invertible $\textit{forward process (FP)}$. Inverse problems can be formalized as follows \cite{ongie2020deep},
\begin{equation}
    \mathbf{y}=A(\mathbf{x})+\mathbf{n}_A,
\end{equation}
where $A$ is the forward operator, an approximate representation of FP, and $\mathbf{n}_A$ is an additive noise. Alternatively, non-additive noise can be considered in a more general model, i.e., $\mathbf{y}=\mathcal{N}(A(\mathbf{x}))$ for the noise distribution $\mathcal{N}(\cdot)$. It is important to note that the FP and the forward operator $A$ are not necessarily identical, in fact, the FP is often a black box and the forward operator is an approximate model of FP. A typical inverse problem can be solved by minimizing the loss function
\begin{equation}
    \text{MSE}(A(\mathbf{x}),\mathbf{y})=\|A(\mathbf{x})-\mathbf{y}\|^2_2
    \label{eq:compressed_sensing}
\end{equation}
with respect to $\mathbf{x}$ with the help of prior assumptions on $\mathbf{x}$.

%%%%%%%%%%%%%%%%%%%%%%%%%%%%%%%%%%%%%%%%%%%%%%
\begin{figure}[!t]
\centering
\includegraphics[width=3.4in]{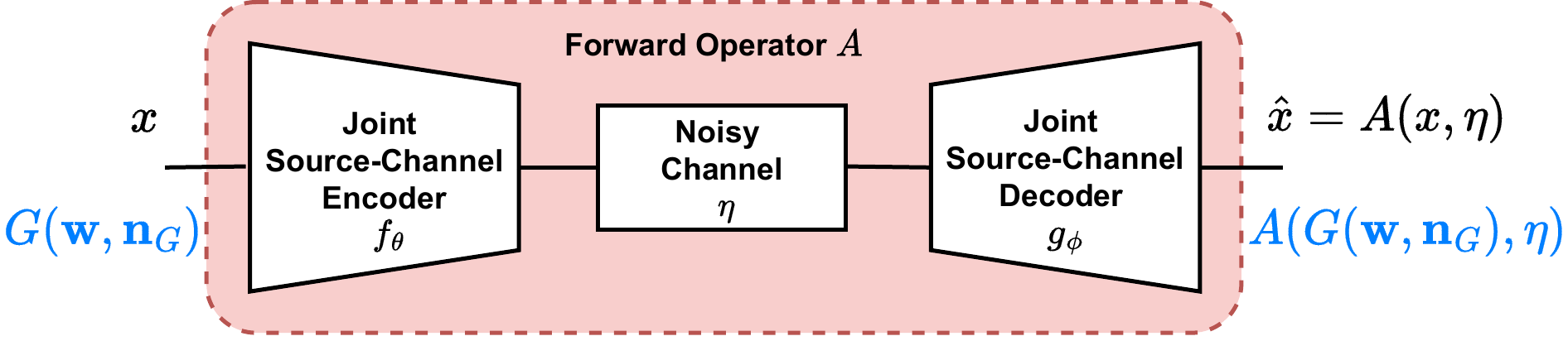}
\caption{DeepJSCC model as the forward operator $A$, where $\eta$ is the AWGN channel function, $G(\cdot)$ is the generator network of a GAN, and $\mathbf{w}$ is the latent vector input to the generator.}
\label{fig:ForwardModel}
\end{figure}
%%%%%%%%%%%%%%%%%%%%%%%%%%%%%%%%%%%%%%%%%%%%%%%

In our communication setting, the source signal $\mathbf{x}$ goes through the encoder, channel, and decoder of the SotA DeepJSCC model in \cite{tung2021deepwive}. We assume that the FP is only one realization of the DeepJSCC model. We have the knowledge of channel statistics, but not the noise realizations. Therefore, the FP is only partially known due to the stochastic channel noise. Our goal is to invert the FP and reconstruct the source signal $\mathbf{x}$ from the observations $\mathbf{\hat{x}}$ with high perceptual quality. We model the encoder and the decoder functions of the DeepJSCC model \cite{tung2021deepwive}, i.e., $f_{\theta}$ and $g_{\phi}$, as DNNs parameterized by $\theta$ and $\phi$, respectively. Then, 
we represent our forward operator $A$ as a non-linear process with AWGN noise function $\eta$, i.e.,
\begin{align}
    \mathbf{y}  &= A(\mathbf{x},\eta) \\
                &= g_{\phi}(\eta(f_{\theta}(\mathbf{x}),\sigma^2)) \\
                &= \mathbf{\hat{x}}.
\end{align}

Extreme physical conditions, such as very low BCR or very low SNR in the communication pipeline introduce high distortion in the signal reconstructed by DeepJSCC, $\mathbf{\hat{x}}$. This might lead to significant loss of semantic similarity between the original signal and the reconstruction, especially when the source data distribution is complex and high dimensional (e.g., high-quality images or videos). In classical signal reconstruction methods, some prior knowledge about the properties of $\mathbf{x}$, such as sparsity, dictionary, or geometric properties, are used to minimize the loss (\ref{eq:compressed_sensing}) effectively. Here, our assumption is that the distribution of our input signal, i.e., images from a certain dataset, is in the range of a GAN generator function $G: \mathbb{R}^{q}\rightarrow \mathbb{R}^m$, which is parameterized by DNNs trained on the same context as the input distribution. In general, DGMs have recently demonstrated unprecedented visual results for image generation and have been shown to represent complex-high dimensional distributions in inverse problems successfully \cite{ongie2020deep,bora2017compressed,daras2021intermediate}. Primary examples for DGMs are VAEs \cite{kingma2013auto}, Diffusion Models \cite{ho2020denoising} and GANs \cite{goodfellow2014generative}. In this paper, we use the StyleGAN-2 \cite{Karras2019stylegan2} generator in the InverseJSCC approach to the semantic communication problem defined in Section \ref{sec:problem_statement}. 

In InverseJSCC, we are inspired by a general inverse problem approach called intermediate layer optimization (ILO) \cite{daras2021intermediate}, which utilizes a pre-trained StyleGAN-2 generator for various inverse problems. Although we are mainly interested in high-quality face images, the same ideas can be applied to other domains. Here, ILO solves an inverse problem objective by adaptively changing the StyleGAN-2 layer to be optimized, moving from the initial latent vector to intermediate layers closer to the output. We solve a slightly modified optimization problem:

\begin{align}
    \min \limits_{\mathbf{w},\mathbf{n}_G} \ & \lambda_1\|A(G(\mathbf{w},\mathbf{n}_G),\eta)-\mathbf{\hat{x}}\|^2_2+
    \lambda_2 \mathcal{L}(A(G(\mathbf{w},\mathbf{n}_G),\eta),\mathbf{\hat{x}}) \nonumber \\
    & + \lambda_3 \mathcal{R}(\mathbf{w},\mathbf{\hat{x}},G,\mathbf{n}_G),
    \label{eq:ILO}
\end{align}
where $G(\cdot)$ is the pre-trained generator network, and $\mathbf{w}$ and $\mathbf{n}_G$ are the latent input and the noise vector of the generator, respectively. While the latent vector controls the style parts of the generated image, the noise fed to the generator determines the high-resolution details. $\mathcal{L}(\cdot)$ is the LPIPS loss function that uses VGG \cite{zhang2018unreasonable}, and $\mathcal{R}$ is the regularization loss defined by 
\begin{equation}
\mathcal{R}=\lambda_4 \mathcal{L}(G(\mathbf{w},\mathbf{n}_G),\mathbf{\hat{x}}) + \lambda_5 \text{GEO}(\mathbf{w}).
\end{equation}
Here, $\text{GEO}(\cdot)$ is the geodesic distance which represents the shortest path between two points on a curved surface. This term regularizes the deviation of the latent vector $\mathbf{w}$. Unlike in ILO, we use an additional LPIPS regularization term that helps make sure that the generator output does not divert far from the reconstructed input perceptually. 

\begin{figure}[pt]
\centering
\includegraphics[width=3.4in]{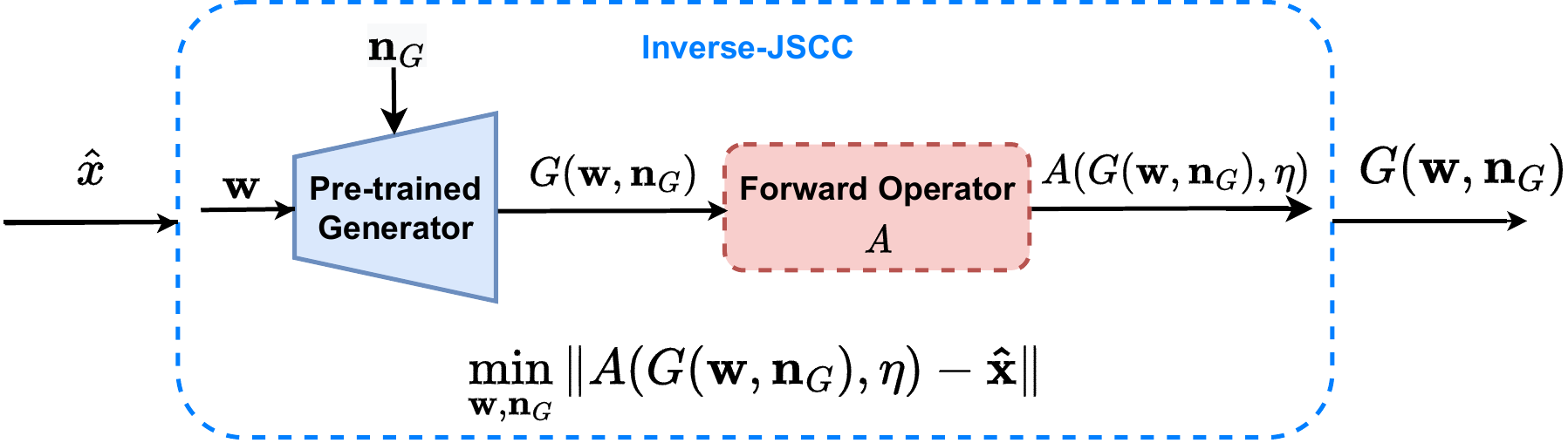}%
\caption{InverseJSCC scheme applied to the signal reconstructed by the DeepJSCC encoder/decoder pair in an unsupervised way.}\label{fig:InverseJSCC}
\end{figure}

Recall that the forward operator $A$ is the DeepJSCC model in Fig. \ref{fig:ForwardModel}, which includes a pre-trained encoder/decoder pair $(f_\theta, g_\phi)$ and the channel model $\eta(\cdot,\sigma^2)$, and is a partial description of the FP due to the stochasticity of the channel.
Fig. \ref{fig:InverseJSCC} shows the InverseJSCC scheme that reconstructs $\mathbf{x}$ from its distorted measurements $\mathbf{\hat{x}}$ with the help of the pre-trained generator.
The forward operator $A$ can be trained on any dataset regardless of the source dataset, whereas StyleGAN-2 generator $G$ is pre-trained on the same context data as the source dataset. 
The generator $G$ can be decomposed into multiple layers as
\begin{equation}
 G=G_4\circ G_3 \circ G_2 \circ G_1, \text{ where }
    \begin{cases}
      G_1: & \mathbb{R}^q\rightarrow\mathbb{R}^{t_1}\\
      G_2: & \mathbb{R}^{t_1}\rightarrow\mathbb{R}^{t_2}\\
      G_3: & \mathbb{R}^{t_2}\rightarrow\mathbb{R}^{t_3}\\
      G_4: & \mathbb{R}^{t_3}\rightarrow\mathbb{R}^{m}.
    \end{cases}       
\end{equation}

%%%%%%%%%%%%%%%%%%%%%%%%%%%%%%%%%%%%%%%%%%%%%%%%%
%%%%%%%%%%%%%%%%%%%%%%%%%%%%%%%%%%%%%%%%%%%%%%%%%

Similarly to ILO, the optimization in (\ref{eq:ILO}) is performed over $\mathbf{w}$ and $\mathbf{n}_G$. The optimized latent vector $\mathbf{w}^*$, which is the input to the first layer of the generator $G_1$, is used for the intermediate latent representation of the input to $G_2$, i.e., $\mathbf{\hat{w}} = G_1(\mathbf{w}^*)$. This recursive optimization continues until the intermediate latent inputs are optimized for all the intended number of layers. The noise vector $\mathbf{n}_G$, on the other hand, is optimized up to a pre-determined layer. Note that the weights of the pre-trained $A$ and $G$ functions are fixed, and only the latent input and noise vectors are optimized.

\begin{figure*}[!t]
\centering
\hspace{-7mm} \subfloat[Joint source-channel encoder $f_{\theta}$]{\includegraphics[width=7in]{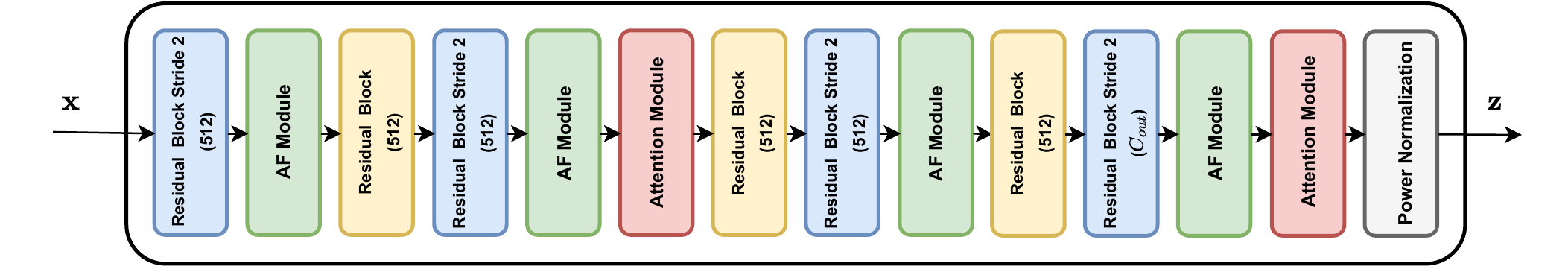}%
\label{fig:encoder_arch_inverse}}
\vfil
\subfloat[Joint source-channel decoder $g_{\phi}$]{\includegraphics[width=7.1in]{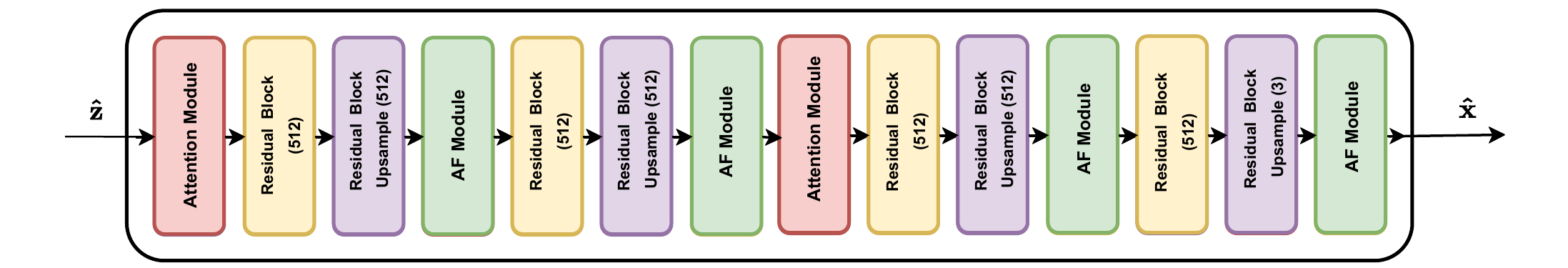}%
\label{fig:decoder_arch_inverse}}
\caption{DNN architecture of (a) the encoder and (b) the decoder of the forward operator $A$.}
\label{fig:architecture_inverse}
\end{figure*}

InverseJSCC has the same encoder/decoder structure used for the key frame structure in \cite{tung2021deepwive}, which is shown in Fig. \ref{fig:architecture_inverse}. We also incorporate the same residual blocks, attention modules, and attention feature (AF) modules proposed in \cite{xu2021wireless}. AF module allows training a single model that works for a range of SNRs by randomly choosing the channel SNR during training and providing the AF modules with the current SNR. A single model trained with AF modules is shown by \cite{xu2021wireless} to perform as well as the models trained for each SNR individually. Moreover, StyleGAN-2 generator structure and the recursive optimization of the intermediate latent vectors are the same as ILO \cite{daras2021intermediate}. We apply our own objective as well as the forward operator as shown in Eqn. (\ref{eq:ILO}) and Fig. \ref{fig:InverseJSCC}.

Our forward operator $A$ lies within the class of non-linear and stochastic forward operators in the inverse problem taxonomy \cite{ongie2020deep}, which is only partially known during test time due to the stochastic noise function $\eta$. The novelty of our InverseJSCC approach is that inverse problem solution has not been considered in the context of wireless communication, and partially known forward operators have not been investigated in this context yet. In the numerical results section, we show for the image domain that InverseJSCC provides perceptually high-quality images that better preserve the semantics of the source image despite communication over the noisy channel. This is achieved by exploiting the remarkable representation quality that DGMs have achieved in the image domain in the past few years. In particular, the improvements that InverseJSCC provides compared to DeepJSCC become more significant as the physical condition of the channel deteriorates. 
Moreover, we also show in Section \ref{sec:numerical_results} that InverseJSCC allows transferability by generating high-quality images even though the forward operator is trained on a different dataset. This is due to the fact that the inverse problem approach is based on successfully inverting the FP with the help of function $A$, which gives the flexibility to use various forward operators.

\subsection{End-to-end Semantic Communication}
\label{sec:end_to_end}

InverseJSCC, presented in the previous section, improves the perceptual quality of a pre-trained DeepJSCC by exploiting the generative capability of a pre-trained GAN.
Moreover, it allows flexibility to use encoder/decoder pairs trained on datasets that are not the test set, since it solves an inverse problem, which is only sensitive to the forward operator $A$. On the other hand, in a scenario where we can train the whole encoder/decoder structure with the same dataset as the test dataset, the best approach would indeed be training an encoder/decoder pair in a supervised way. While InverseJSCC solves an optimization problem over the layers of the generator network in test time, the alternative GenerativeJSCC approach involves an end-to-end optimization during training, not testing.

Similarly to InverseJSCC, our goal in the end-to-end design is to maximize the semantic similarity of the reconstructed signal to the source image while also retaining its perceptual quality in the presence of extreme channel conditions. Accordingly, the GenerativeJSCC scheme is an end-to-end strategy based on jointly training the encoder and the decoder functions represented by DNNs parameterized by $\theta$ and $\psi$. The source image $\mathbf{x}$ is sent to the encoding function $f_{\theta}$, which has the same structure as shown previously in Fig. \ref{fig:encoder_arch_inverse}, and normalized by (\ref{eq:power_normalization}) at the last layer to satisfy the power constraint. Then encoded signal $\mathbf{z}$ is transmitted over the AWGN channel with the transfer function (\ref{eq:channel}). The receiver observes the corrupted signal $\mathbf{\hat{z}}$, and decodes it into an approximation of the source signal as follows:
\begin{equation}
    \mathbf{\hat{x}}=g_{\psi}(\eta(f_{\theta}(\mathbf{x}), \sigma^2)),
\end{equation}
where the decoding function $g_{\psi}$ is represented by a DNN with parameters $\psi$. As we mentioned in the previous section, AF modules allow end-to-end training of a model that can be adapted for a range of channel SNRs. The current SNR is known by the encoder and the decoder, and it is provided to the AF modules of both models.

In GenerativeJSCC, we employ a weighted distortion metric as follows:
\begin{equation}
    \min\limits_{f_{\theta},g_{\psi}} \gamma_1 \|\mathbf{x}-\mathbf{\hat{x}}\|^2_2
    +\gamma_2  \mathcal{L}(\mathbf{x},\mathbf{\hat{x}}),
    \label{eq:GenerativeJSCC_objective}
\end{equation}
where $\gamma_1$ and $\gamma_2$ are the weights of the MSE and LPIPS loss $\mathcal{L}(\cdot)$.
Here, we impose our requirement to maximize the semantic content of the source at the reconstruction by measuring it with the learned perception, i.e., LPIPS, and optimizing it together with MSE. 
Finally, given the distortion metric defined by Eqn. (\ref{eq:GenerativeJSCC_objective}) between the original source signal and its reconstruction, the network weights $(\theta,\psi)$ are updated via backpropagation with respect to the gradient of the distortion.

\begin{figure*}[!t]
\centering
\includegraphics[width=7.1in]{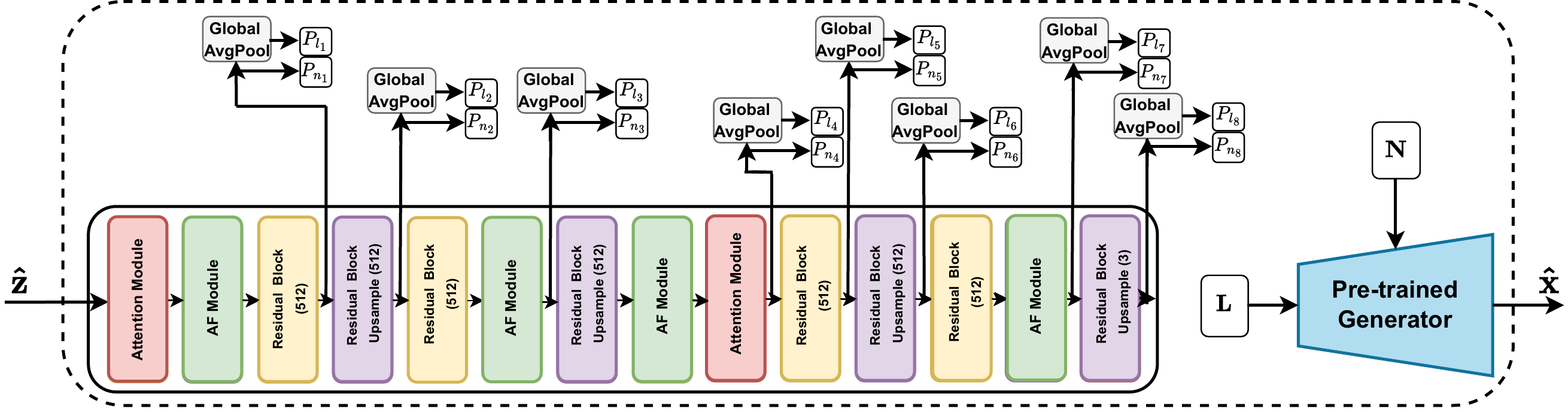}%
\caption{DNN architecture of the decoder function $g_{\psi}$ of GenerativeJSCC.}
\label{fig:e2e_decoder}
\end{figure*}

We focus on the transmission of face images when designing the GenerativeJSCC architecture. While we use the same structure shown in Fig. \ref{fig:encoder_arch_inverse} for the encoder, we propose a novel decoder structure that uses the StyleGAN-2 \cite{Karras2019stylegan2} generator architecture, and residual blocks introduced in \cite{tung2021deepwive} as well as AF modules. 

The GenerativeJSCC decoder contains residual networks and the StyleGAN-2 generator that accepts two types of inputs: (1) the mapped latent vector $\bm{L}$, which determines the style parts of the generated image, and (2) the stochastic noise $\bm{N}$, which is typically drawn from a normal distribution and determines the high-resolution details. We model our decoder such that the noisy signal at the receiver input is transformed into the noise vector $\bm{N}$ and the initial latent vector $\bm{\tilde{L}}$, which is then mapped to $\bm{L}$ by a multi-layer linear network with Leaky ReLU activation. 
Fig. \ref{fig:e2e_decoder} shows the decoder architecture in more detail. While the channel output $\mathbf{\hat{z}}$ goes through the attention modules, AF modules and residual blocks introduced in \cite{tung2021deepwive}; the output of each residual block is split into three parts. The first split is followed by a global average pooling and a projection function $P_{l}$: a $1 \times 1$ convolution that produces parts of the initial latent vector $\bm{\tilde{L}}$. The second branch directly goes to a projection $P_{n}$: a $1 \times 1$ convolution that is used to produce the noise vector in the form of single channel images. The third split is followed by the next residual block. 
Then, the projections are concatenated as $\bm{\tilde{L}}=\text{Concat}\{P_{l_1},P_{l_2}, \dots, P_{l_8}\}$ and as $\bm{N}=\text{Concat}\{P_{n_1},P_{n_2}, \dots, P_{n_8}\}$, where subscripts represent the order of the intermediate layer outputs. Finally, the initial latent vector is mapped to the latent vector $\bm{L}$, and it is fed to the generator together with the noise vector to generate high-resolution face images.
Note that the weights of the StyleGAN-2 generator are fixed during the training of the encoder/decoder pair of GenerativeJSCC.

We use a two-stage training scheme for learning the latent vector and the noise to maximize the perceptual quality of the reconstructions. The first stage only involves the reconstruction from the latent vector and disables learning of the noise maps. The second stage fine-tunes the model by training the layers learning the noise maps, which improves the details of the generated images. We implement this strategy by end-to-end training the model with the same learning rate for the $P_{\ell}$ layers and the backbone of the model, whereas we do not train the $P_n$ layers initially. In the second stage, we train all layers of the model but we adjust the learning rate of the $P_n$ layers to the initial learning rate of the backbone, whereas the learning rate for the backbone and $P_{\ell}$ is divided by $100$.

\section{Numerical Results}
\label{sec:numerical_results}
In this section, we present our experimental results to demonstrate and compare the performance of InverseJSCC and GenerativeJSCC. Both schemes are tested over an AWGN channel with the transfer function (\ref{eq:channel}), while the generator network refers to StyleGAN-2 generator which is pre-trained on the Flickr-Faces-HQ (FFHQ) dataset of high-quality human face images. We also use $512 \times 512$ CelebA-HQ dataset \cite{karras2017progressive}, which consists of $30000$ high-quality celebrity images, and a subset of ImageNet dataset with $1050000$ samples of size $256 \times 256$. Both datasets are split as $8:1:1$ for training, validation, and testing, respectively.

\subsection{Performance Metrics}
We consider various metrics to measure the distortion between the generated images and the input image throughout the numerical results: peak signal-to-noise-ratio (PSNR), multi-scale structural similarity
index measure (MS-SSIM) and LPIPS. They are defined as follows:
\begin{equation}
   \hspace{-1.6cm} \text{PSNR}(\mathbf{x},\mathbf{\hat{x}})=10\log_{10}\Big( \frac{255^2}{\text{MSE}(\mathbf{x,\hat{x}})}\Big) \text{~dB}.
\end{equation}
\begin{align}
& \hspace{-0.6cm} \text{MS-SSIM}(\mathbf{x,\hat{x}}) \nonumber \\
&=[l_M(\mathbf{x,\hat{x}})]^{\alpha_M}\prod\limits_{j=1}^{M}[c_j(\mathbf{x,\hat{x}})]^{\beta_j}[s_j(\mathbf{x,\hat{x}})]^{\gamma_j},
\end{align}
where
\begin{align}
l_M(\mathbf{x,\hat{x}}) = \frac{2\mu_{\mathbf{x}}\mu_{\mathbf{\hat{x}}}+c_1}{\mu_{\mathbf{x}}^2+\mu_{\mathbf{\hat{x}}}^2+c_1}, \\
c_j(\mathbf{x,\hat{x}}) = \frac{2\sigma_{\mathbf{x}}\sigma_{\mathbf{\hat{x}}}+c_2}{\sigma_{\mathbf{x}}^2+\sigma_{\mathbf{\hat{x}}}^2+c_2}, \\
s_j(\mathbf{x,\hat{x}}) = \frac{\sigma_{\mathbf{x}\mathbf{\hat{x}}}+c_3}{\sigma_{\mathbf{x}}+\sigma_{\mathbf{\hat{x}}}+c_3}.
\end{align}
Here, $\mu_{\mathbf{x}}, \sigma_{\mathbf{x}}^2, \sigma_{\mathbf{x,\hat{x}}}^2$ are the mean and variance of $\mathbf{x}$, and the covariance between $\mathbf{x}$ and $\mathbf{\hat{x}}$, respectively. Moreover, $\alpha_M, \beta_j$ and $\gamma_j$ are the weights of the components; while $c_1, c_2$ and $c_3$ are coefficients for numeric stability. Subscript $j$ represents different downsampling scale of $\{\mathbf{x},\mathbf{\hat{x}}\}$. Similar to \cite{tung2021deepwive}, we use the default parameters for $(\alpha_M,\beta_j,\gamma_j)$ in the original paper \cite{wang2003multiscale}. MS-SSIM has been shown to approximate human visual perception well on various image and video databases.

LPIPS loss was proposed to measure the perceptual similarity loss between two images \cite{zhang2018unreasonable}, and it has been shown to match human perception well. It essentially computes the similarity between the activations of two image patches for some pre-defined network, such as VGG or AlexNet. Note that, a lower LPIPS score is better since it means that image patches are perceptually more similar.

\subsection{InverseJSCC Results}
\label{sec:Inverse-Results}
We consider an edge case in which the channel SNR is in the range $\{-5,-4, \dots, 4, 5\}$ dB and the BCR is $\rho=\{0.0013, 0.0052\}$. According to the scenario defined in Section \ref{sec:inverse_problems}, we have a DeepJSCC model which reconstructs source images with high distortion due to extreme physical conditions. Note that all DeepJSCC models used as the forward operators are trained by optimizing the objective $d(\mathbf{x},\mathbf{\hat{x}})=\text{MSE}(\mathbf{x},\mathbf{\hat{x}})+\text{LPIPS}(\mathbf{x},\mathbf{\hat{x}})$. Our goal is to achieve a better perception quality in the reconstructed images without sacrificing the distortion performance with respect to the existing alternatives. Here, we follow an unsupervised approach and apply InverseJSCC to the distorted reconstructions at the output of the DeepJSCC decoder.

\begin{figure}[pt]
\centering
\includegraphics[width=3.4in,height=3in]{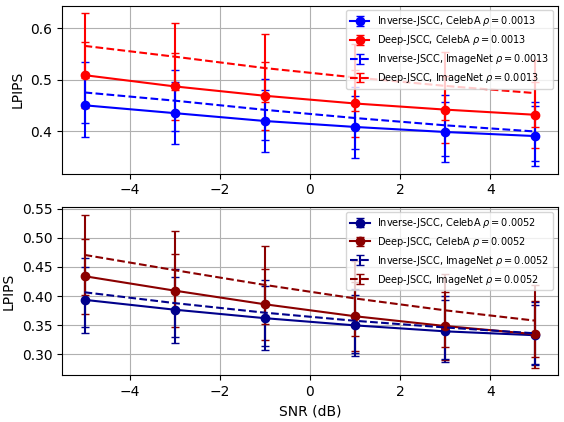}%
\caption{LPIPS loss between the source and the reconstructed image w.r.t. channel SNR for DeepJSCC (trained with CelebA-HQ), InverseJSCC ($A$ is trained with CelebA-HQ), DeepJSCC (trained with ImageNet) and InverseJSCC ($A$ is trained with ImageNet) for $\rho=\{0.0013,0.0052\}$.}
\label{fig:performanceInverse}
\end{figure}

We represent both the FP and the forward operator $A$ of InverseJSCC with the SotA DeepJSCC architecture shown in Fig. \ref{fig:architecture_inverse}. However, they are not identical due to the stochastic channel noise. We perform experiments for two major cases:
First, each encoder/decoder pair representing the FP and $A$ is pre-trained with CelebA-HQ dataset for the given $\rho$ and SNR values, and the InverseJSCC is tested against the same dataset. 
In the second case, the FP and the forward  operator $A$ are both pre-trained with ImageNet dataset, and the InverseJSCC is tested against the CelebA-HQ dataset.

All DeepJSCC models are trained for $\text{SNR}_{\text{Train}} \in \{-5,-4, \dots, 4, 5\}$ dB using PyTorch framework \cite{paszke2017automatic}. To train the encoder and decoder networks of the FP and the forward operator $A$, we utilize Adam optimizer \cite{kingma2013auto} with $\beta_1=0.9$, $\beta_2=0.99$ and an initial learning rate $l_r=0.0001$, which is multiplied by a factor $0.8$ when the validation error does not improve in 4 epochs. We also set $\lambda_1=1, \lambda_2=1, \lambda=1$ and $\lambda_5=0.01$ as given in ILO, and we selected $\lambda_4=0.1$.

\begin{figure*}[!t]
\centering
\subfloat[Original images from the CelebA-HQ dataset.]{\includegraphics[width=7in]{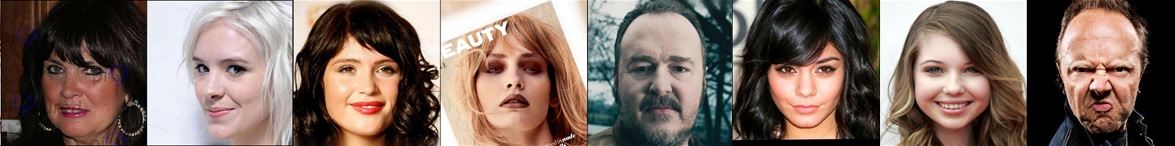}\label{fig:Inv_original}}\\
\includegraphics[width=7in]{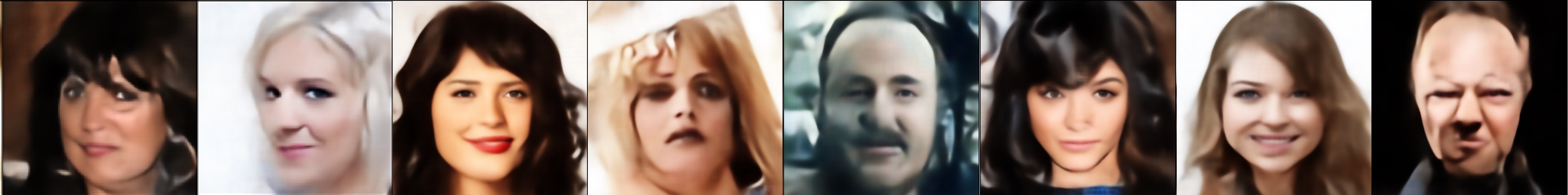}\\
\includegraphics[width=7in]{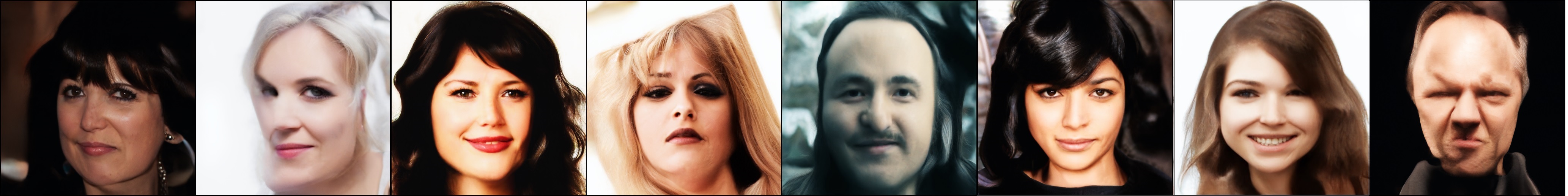}\\
\includegraphics[width=7in]{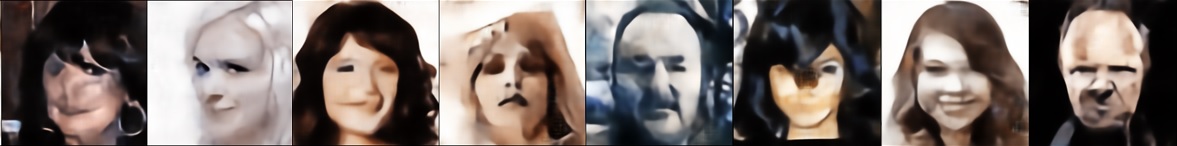}\\
\vspace{-0.33cm}
\subfloat[First row: Reconstructions by DeepJSCC(CelebA-HQ), Second row: InverseJSCC(CelebA-HQ), Third row: DeepJSCC(ImageNet) and Fourth row: InverseJSCC(ImageNet), respectively, for SNR$=1$.]{
\hspace{-0.25cm}
\includegraphics[width=7in]{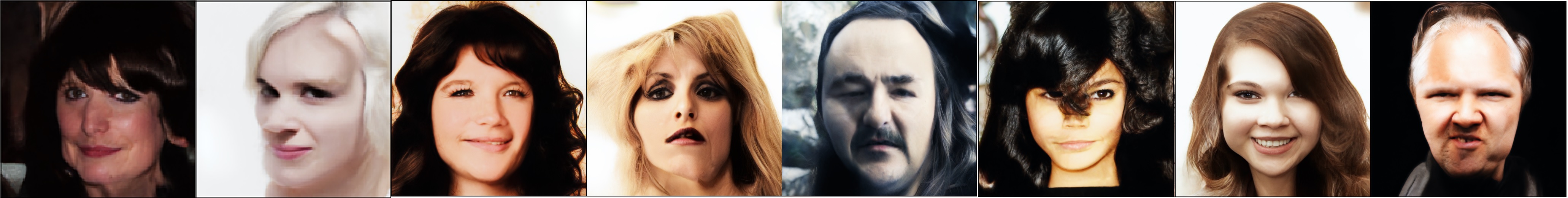}%
\label{fig:Inv_SNR1}}\\
% \includegraphics[width=7in]{figures/InverseJSCC_ImageNet_snr3_cout2.jpg}\\
% \vspace{-0.32cm}
% \subfloat[CelebA images reconstructed by DeepJSCC and InverseJSCC, respectively, for $SNR=3$.]{
% \hspace{-0.25cm}
% \includegraphics[width=7in]{figures/InverseJSCC_ImageNet_ILO_snr3_cout2.jpg}%
% \label{fig:Inv_SNR3}}\\
\includegraphics[width=7in]{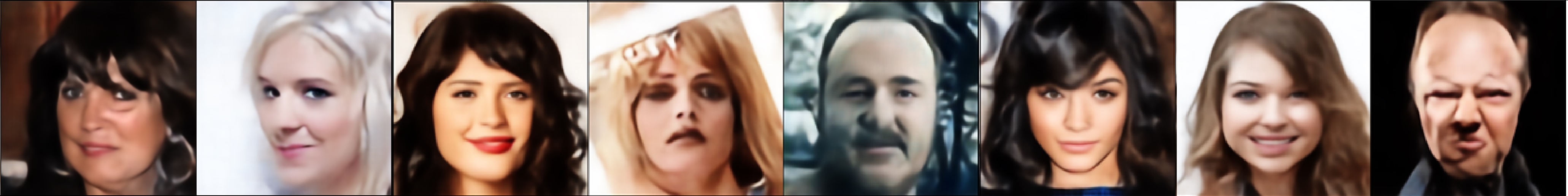}\\
\includegraphics[width=7in]{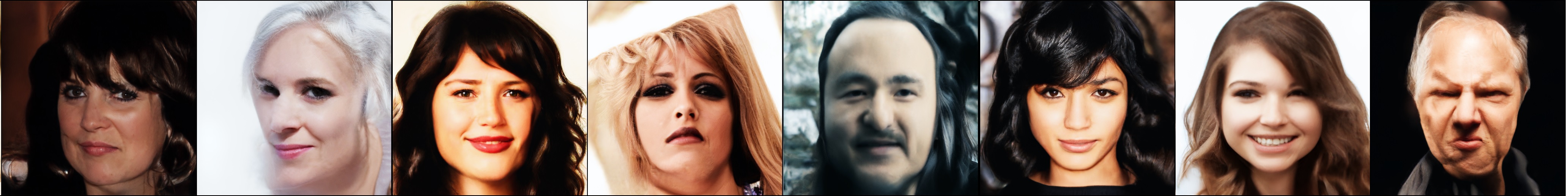}\\
\includegraphics[width=7in]{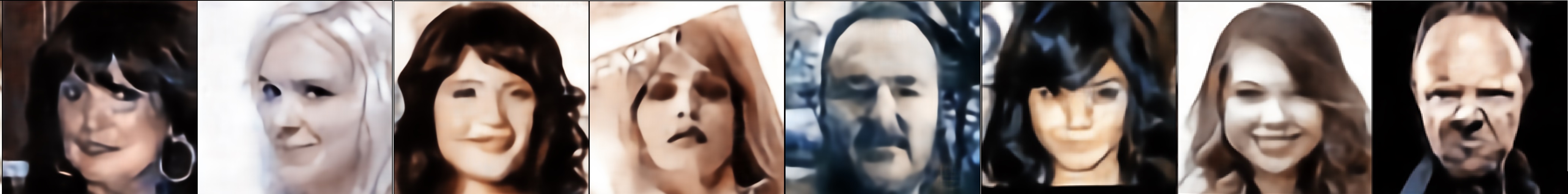}\\
\vspace{-0.33cm}
\subfloat[First row: Reconstructions by DeepJSCC(CelebA-HQ), Second row: InverseJSCC(CelebA-HQ), Third row: DeepJSCC(ImageNet) and Fourth row: InverseJSCC(ImageNet), respectively, for SNR$=5$.]{
\hspace{-0.25cm}
\includegraphics[width=7in]{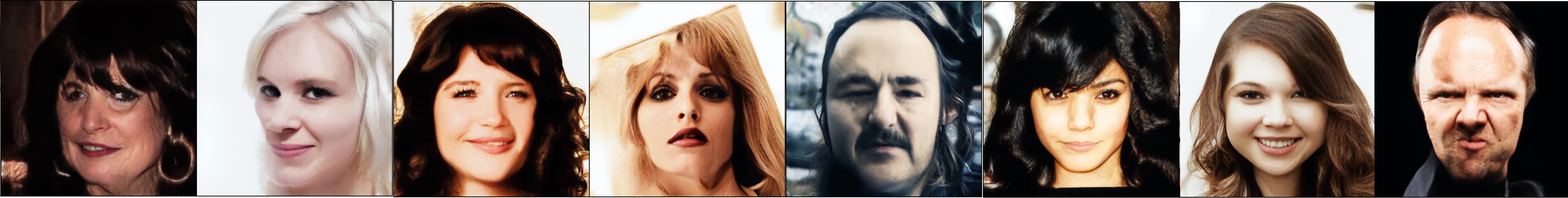}%
\label{fig:Inv_SNR5}}
\caption{Original and reconstructed CelebA-HQ images, when DeepJSCC model and the forward operator $A$ of InverseJSCC are trained on ImageNet for $\rho=0.0013$ and $SNR=\{1,3,5\}$. Inverse problem solution in InverseJSCC allows $A$ trained with a different dataset. }
\label{fig:ReconstructionsInverse}
\end{figure*}

Fig. \ref{fig:performanceInverse} shows the performances of DeepJSCC and InverseJSCC models in terms of LPIPS loss with respect to channel SNR (dB). The top plot depicts the results for $\rho=0.0013$ and the bottom for $\rho=0.0052$. ``DeepJSCC, CelebA-HQ'' and ``DeepJSCC, ImageNet'' represent the encoder/decoder pairs with the structure shown in Fig. \ref{fig:architecture_inverse} and trained with CelebA-HQ and ImageNet datasets, respectively. 
``InverseJSCC, CelebA-HQ'' and ``InverseJSCC, ImageNet'' represent our proposed method described in Fig. \ref{fig:InverseJSCC} with the forward operators trained with CelebA-HQ and ImageNet datasets, respectively. 
These models represent the FP's and $A$'s of the two major cases we described above. In Fig. \ref{fig:performanceInverse}, smaller LPIPS loss implies better perceptual similarity between the ground-truth and the generated images in the presented schemes. As expected, we observe better LPIPS score for both DeepJSCC and InverseJSCC models when the encoder/decoder networks involved are trained and tested with the CelebA-HQ dataset, and worse when they are trained in another domain, i.e., ImageNet, but tested with CelebA-HQ. However, in both cases, InverseJSCC improves the perceptual similarity (measured by LPIPS) of the reconstructed images by the DeepJSCC model significantly, even when there is a domain mismatch.

The noteworthy effect of InverseJSCC on the reconstructed images can be seen in Fig. \ref{fig:ReconstructionsInverse} clearly. Fig. \ref{fig:Inv_original} shows the ground-truth images from CelebA-HQ test-set; Fig.\ref{fig:Inv_SNR1} and Fig.\ref{fig:Inv_SNR5} both show the reconstructed images at the output of DeepJSCC (trained with CelebA-HQ), InverseJSCC ($A$ is trained with CelebA-HQ), DeepJSCC (trained with ImageNet) and InverseJSCC ($A$ is trained with ImageNet) from top to bottom when the channel SNR is $1~$dB and $5~$dB, respectively. The facial details introduced by InverseJSCC crucially improve the perceptual quality of the images reconstructed by DeepJSCC. Hence, the generated images look as realistic as the ground truth. Moreover, InverseJSCC generates considerably higher quality face images even when the available encoder/decoder networks are trained on a different domain, i.e., ImageNet. This can be attributed to two reasons: (1) InverseJSCC inverts the FP regardless of its training set, and (2) the source distribution is in the range of the StyleGAN-2 generator, therefore, the $G(\cdot)$ function is able to generate realistic-looking face images that also look similar to the ground-truth. 

\begin{figure}[!t]
\centering
\subfloat[LPIPS loss versus SNR.]{\includegraphics[width=3.4in]{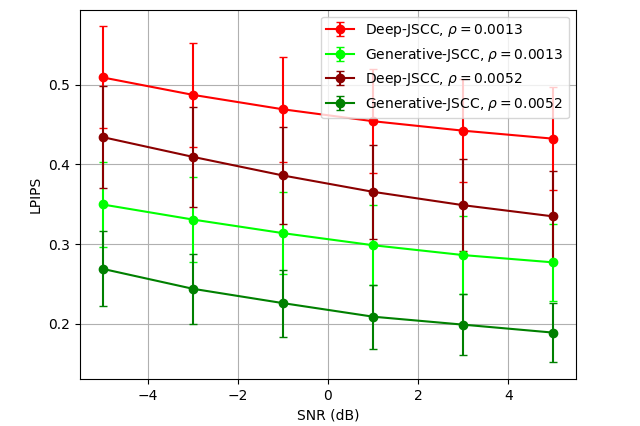}%
\label{fig:lpips}}\\
\vspace{-4mm}
\subfloat[MS-SSIM versus SNR.]{\includegraphics[width=3.4in]{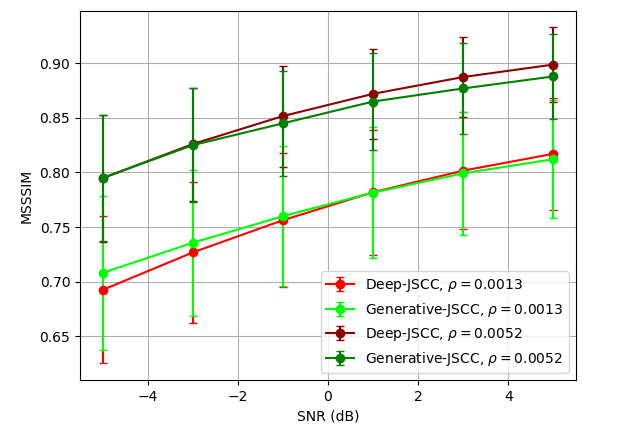}%
\label{fig:msssim}}\\
\vspace{-4mm}
\subfloat[PSNR versus SNR.]{\includegraphics[width=3.4in]{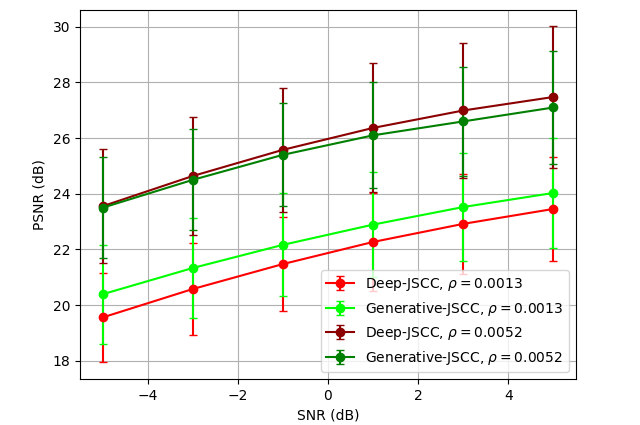}%
\label{fig:psnr}}
\caption{Performance comparison between DeepJSCC and GenerativeJSCC in terms of (a) LPIPS, (b) MS-SSIM and (c) PSNR metrics w.r.t channel SNR for $\rho=\{0.0013,0.0052\}$.}
\label{fig:performanceGenerative}
\vspace{-0.5cm}
\end{figure}

\subsection{GenerativeJSCC Results}
\label{sec:Generative-Results}
Here again, we consider the channel with SNR = $\{-5,-4, \dots, 4, 5\}$ dB and the BCR of $\rho=\{0.0013,0.0052\}$. We jointly train the encoder, $f_{\theta}$, and the decoder, $g_{\psi}$, of the GenerativeJSCC model with CelebA-HQ dataset by optimizing the weighted sum of MSE and LPIPS losses between the ground-truth and the generated images (\ref{eq:GenerativeJSCC_objective}). As before, we train the networks for a range of SNR utilizing the attention modules, i.e., $\text{SNR}_{Train}\in \{-5, -4, \dots , 4, 5\}$ dB, using PyTorch and Adam optimizer with the same parameters as in Section \ref{sec:Inverse-Results}.

Fig. \ref{fig:performanceGenerative} shows the comparison between DeepJSCC and GenerativeJSCC in terms of average LPIPS loss, MS-SSIM and PSNR with respect to channel SNR (dB). We observe that, for lower BCR, GenerativeJSCC outperforms DeepJSCC not only in terms of perceptual quality but also in terms of pixel-wise distortion. Considering that both models are trained to optimize the same objective given in (\ref{eq:GenerativeJSCC_objective}), this shows the superior reconstructing capabilities of GenerativeJSCC through our novel decoder architecture. On the other hand, GenerativeJSCC outperforms DeepJSCC only in terms of perceptual similarity loss LPIPS for in the larger BCR regime. This is in line with our claim that generative models in JSCC provide higher perceptual quality and thus, more semantic similarity, than pixel-wise similarity in edge cases.

\begin{figure*}[!t]
\centering
\subfloat[Original images from the CelebA-HQ dataset.]{\includegraphics[width=3.5in]{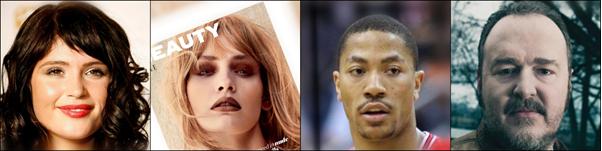}\includegraphics[width=3.5in]{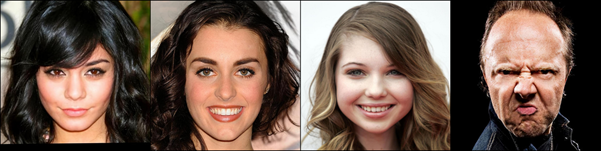}\label{fig:Gen_original}}\\
\includegraphics[width=3.5in]{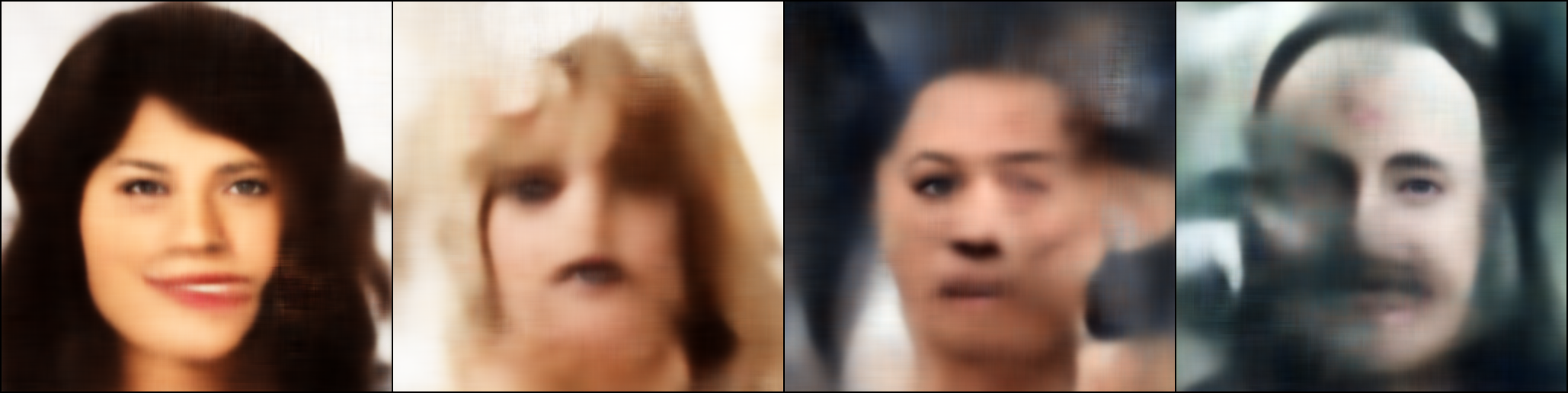}\includegraphics[width=3.5in]{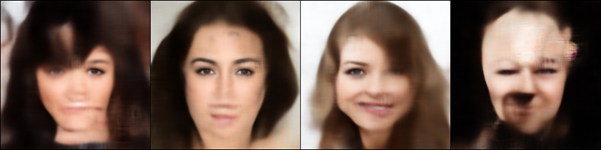}\\
\subfloat[Reconstructions by DeepJSCC and GenerativeJSCC, respectively, for SNR$=-5$.]{\includegraphics[width=3.5in]{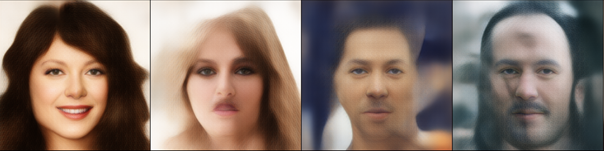}\includegraphics[width=3.5in]{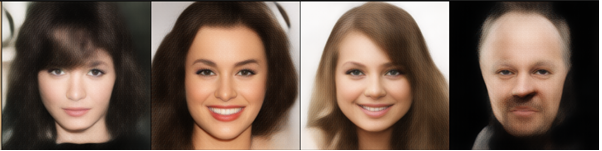} \label{fig:Gen_SNR-5}}\\
\includegraphics[width=3.5in]{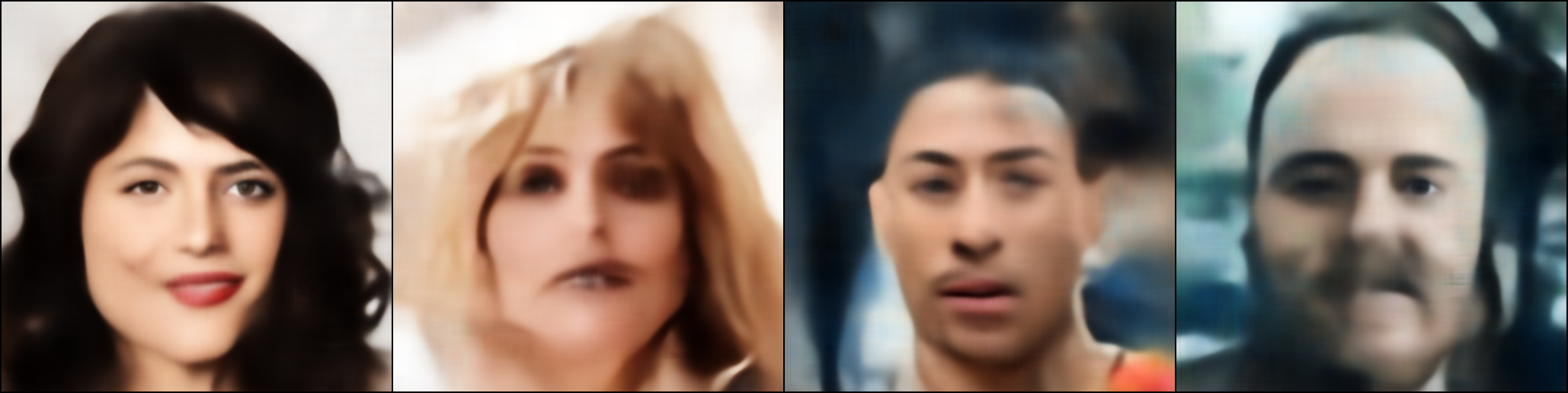}\includegraphics[width=3.5in]{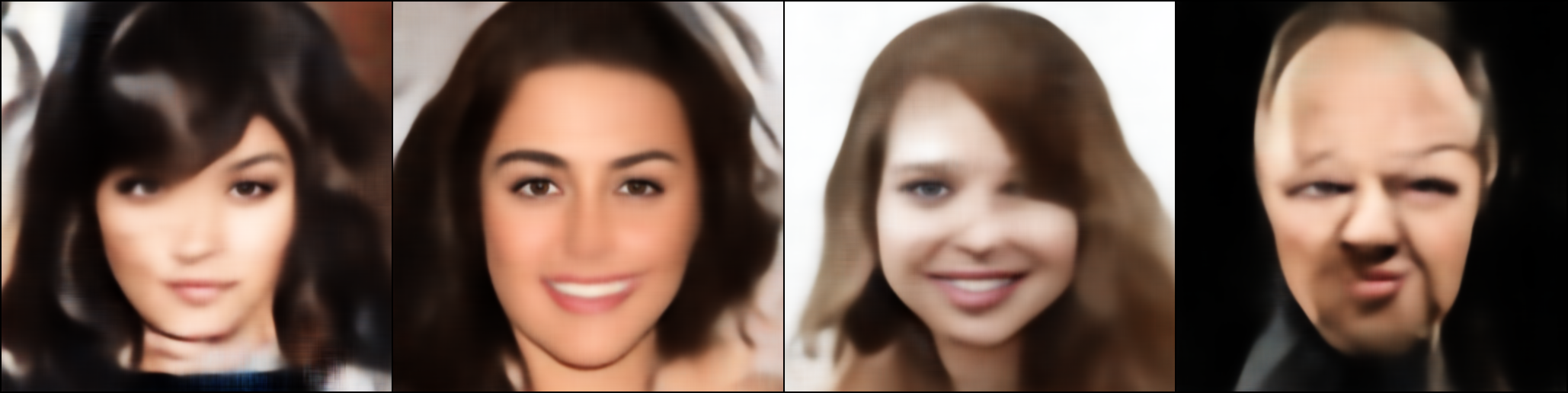}\\
\subfloat[Reconstructions by DeepJSCC and GenerativeJSCC, respectively, for SNR$=-1$.]{\includegraphics[width=3.5in]{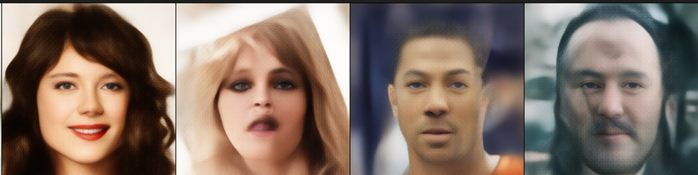}\includegraphics[width=3.5in]{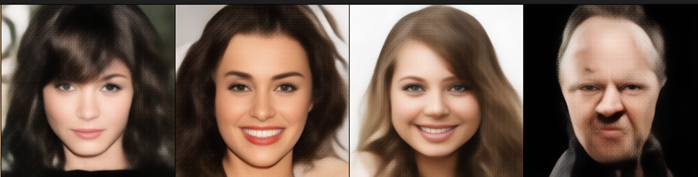} \label{fig:Gen_SNR-1}}\\
\includegraphics[width=3.5in]{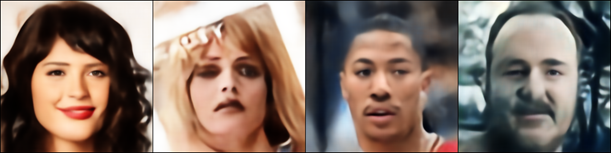}\includegraphics[width=3.5in]{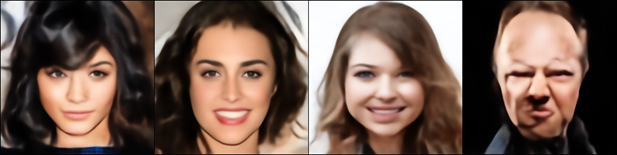}\\
\subfloat[Reconstructions by DeepJSCC and GenerativeJSCC, respectively, for SNR$=5$.]{\includegraphics[width=3.5in]{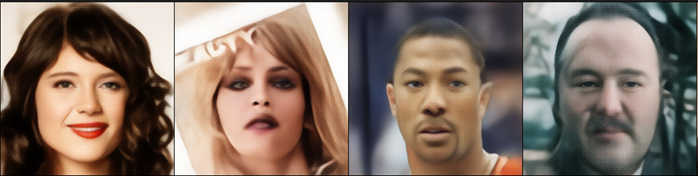}\includegraphics[width=3.5in]{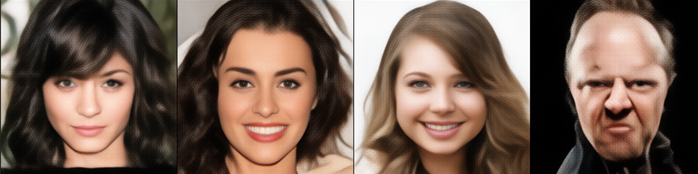} \label{fig:Gen_SNR5}}\\
\caption{Original and reconstructed CelebA-HQ images by DeepJSCC and GenerativeJSCC models for $\rho=0.0013$ and $SNR=\{-5,-1,5\}$. }
\label{fig:ReconstructionsGenerative}
\end{figure*}

We show a visual comparison between the reconstructed images from both models in Fig. \ref{fig:ReconstructionsGenerative}. When the channel SNR is as low as $-5~$dB, the performance difference between DeepJSCC and GenerativeJSCC is much more evident. This shows that the GAN-based decoder of GenerativeJSCC generates higher quality images for human perception, whereas DeepJSCC output distorts facial attributes significantly. As the channel SNR increases to $5~$dB, the image quality difference between two models becomes less obvious, however, GenerativeJSCC still captures the image color and face details impressively better than DeepJSCC.

\section{Conclusion}
In this paper, we presented two DL-based JSCC schemes that incorporate SotA DGMs to improve the perceptual quality of reconstructed images. Both approaches are proposed to tackle edge cases such as low bandwidth and low channel SNR, in which the conventional DeepJSCC scheme typically results in reconstructed images with poor perceptual quality. Our first scheme, InverseJSCC, takes a novel inverse problem approach to the wireless image transmission problem, and improves the classical DeepJSCC scheme by recovering the source image with the help of a StyleGAN-2 generator. It maximizes the semantic similarity between the input and the generated images in terms of the LPIPS metric, which is widely accepted to capture human perception well. Additionally, InverseJSCC performs considerably well when there is a mismatch between the image distributions in training and inference times. Our second approach, GenerativeJSCC, is an end-to-end scheme consisting of an encoder, a non-trainable channel, and a StyleGAN-based decoder. Our results show that GenerativeJSCC outperforms DeepJSCC in terms of perceptual quality metrics learned from human judgments, as well as the distortion metrics used in classical communications.

\bibliographystyle{ieeetr}
\bibliography{refs}

\end{document}